# Stability analysis of the classical ideal gas in nonextensive statistics and the negative specific heat

Liu Liyan, Liu Zhipeng, Guo Lina

*Department of Physics, School of Science, Tianjin University, Tianjin 300072, China*

**Abstract**. We present a stability analysis of the classical ideal gas in a new theory of nonextensive statistics and use the theory to understand the phenomena of negative specific heat in some self-gravitating systems. The stability analysis is made on the basis of the second variation of Tsallis entropy. It is shown that the system is thermodynamically unstable if the nonextensive parameter is $q>5/3$, which is exactly equivalent to the condition of appearance of the negative specific heat.

The ideal gas model, for its simplicity and peculiarity, plays an important role in the celebrated Boltzmann-Gibbs (B-G) statistics. It is on the ground for studying other complex systems from the fields of statistical physics and astrophysics. For example, in the conventional theory of stellar structure [1], the matter of a star is usually assumed to be an ideal gas being in a local equilibrium. In recent years, the theory of nonextensive statistics based on Tsallis entropy has attracted great attention and has proved to be a reasonable generalization of B-G statistics. It is widely applied in the research fields such as self-gravitating systems [2-9], Plasma systems [10-16], astrophysical nuclear reactions [17-20], and magnetic systems [21]. Especially, the successful applications of generalized Boltzmann equation has made it possible from the microscopic dynamics to determine formula expressions of the nonextensive parameter and thereby one can know which systems and under what physical situations the new statistics might be suitable for their statistical description [4,9,11, 22-25]. Due to the basic role of the ideal gas in B-G statistics, it is also of great importance to study its properties in nonextensive statistics. We know that in B-G statistics the classical ideal gas model is studied and applied when assuming the inter-particle week interactions are neglected in the statistical description of the gas. But when using Tsallis $q$-statistics for an ideal gas, we have actually taken into account the week long-range interactions between the particles, although the gas model does not explicitly contain the interaction terms. The effects of the interactions are replaced by introducing the nonextensive parameter $q$ different from unity. For example, we have known now that the nonextensive parameter $q$ can be expressed by the inter-particle interacting potential $\varphi$ as the formulae $k\nabla T + (1-q)m\nabla\varphi = 0$ for the self-gravitating systems [4, 9] and $k\nabla T - (1-q)e\nabla\varphi = 0$ for the plasma systems [11]. Therefore, using Tsallis nonextensive $q$-statistics for an ideal gas is particularly important for the statistical description of some long-range interacting systems, which is different from those physical cases of ideal gas in B-G statistics.

In the nonextensive statistics, it has been found that some properties of an ideal gas, such as the equation of state, the Mayor relation and so on [26, 27], are generally related to the nonextensive parameter $q$. It is well known that the negative specific heat exists in some self-gravitating systems. However, it is difficult to study using the traditional statistics [28]. In other words, the negative nature of the specific heat existing in the classical ideal gas cannot be explained from B-G statistics, but it might be understood by the new theory of nonextensive statistics. In fact, there have been many valuable works using Tsallis $q$-statistics for the classical ideal gas to understand the negative specific heat, such as using the $q$-canonical ensemble theory [29-31], OLM method [32], and the $q$-kinetic theory [33-35]. In this letter, we try to develop a new technique of stability analysis of Tsallis entropy function for the classical ideal gas in the framework of nonextensive statistics and exactly connect the stability with the explanation of the negative specific heat.

The nonextensive statistics is developed based on Tsallis entropy [36] that can be written by

$$S_q = \frac{\int (f^q - f) d^3x d^3v}{1-q}, \quad q \in R \tag{1}$$

where $q$ is a parameter different from unity, used for the description of the degree of nonextensivity of the system under investigation. The basic property of it is nonextensivity or nonadditivity. For example, for two independent systems A and B, the rule of composition for $q \neq 1$ reads:

$$S_q(A+B) = S_q(A) + S_q(B) + (1-q) S_q(A) S_q(B). \tag{2}$$

The extensivity is recovered in the limit $q \to 1$. In this framework, many questions discussed in Boltzmann-Gibbs statistics have been reconsidered. We now introduce the generalized Maxwellian velocity distribution function [37] given by

$$f(\mathbf{v}) = nB_q \left[ 1 - (1-q) \frac{mv^2}{2k_B T} \right]^{1/1-q} \tag{3}$$

where $n$ is the number density of particles of the system, $m$ is the mass of each particle, $T$ is the temperature, $k_B$ is Boltzmann constant and $B_q$ is the normalization constant given by

$$B_q = \begin{cases} \dfrac{1}{4}\left(\dfrac{m}{2\pi k_B T}\right)^{3/2}(1-q)^{1/2}(3-q)(5-3q)\Gamma\left(\dfrac{1}{2}+\dfrac{1}{(1-q)}\right)\bigg/\Gamma\left(\dfrac{1}{(1-q)}\right) & q \le 1 \\ \left(\dfrac{m}{2\pi k_B T}\right)^{3/2}(q-1)^{1/2}\Gamma\left(\dfrac{1}{(q-1)}\right)\bigg/\Gamma\left(-\dfrac{3}{2}+\dfrac{1}{(q-1)}\right) & q > 1 \end{cases}.$$

The total energy of the system [27] can be expressed in the form

$$E = \int \left(\frac{f}{B_0}\right)^q \frac{1}{2}mv^2 d^3x d^3v = \frac{3}{2}Nk_B T\left(\frac{B_q}{B_0}\right)^{q-1}, \tag{4}$$

where $N$ is the total number of the particles and $B_0$ is the dimensional constant. Then the specific heat of the system [27] is obtained of the form as

$$C_V = \frac{5-3q}{2}\frac{3}{2}Nk_B\left(\frac{B_q}{B_0}\right)^{q-1}. \tag{5}$$

We therefore get the conclusions that $C_V > 0$ if $q < 5/3$ and $C_V = 0$ if $q = 5/3$, but $C_V < 0$ if $q > 5/3$, thus leading to the negative specific heat. The negative specific heat is an unusual feature in the traditional theory, usually in astrophysics called the gravothermal catastrophe for the violent instability [28], but seems a usual one in terms of the above *nonextensive* analysis. Here we will show that the phenomena of negative specific heat appearing in the nonextensive statistics is also connected with the instability of the system. For this purpose, we study the stability conditions of a classical idea gas in the nonextensive statistics and then show that the presence of negative specific heat may be consistently explained in terms of instability condition of the system.

We first seek the extreme solution of Tsallis entropy at the fixed mass $M$ and fixed energy $E$ by

$$M = \int mf(\mathbf{v})d^3v d^3x, \tag{6}$$

$$E = \int f^q(\mathbf{v})\frac{1}{2}mv^2 d^3v d^3x, \tag{7}$$

Namely, we let $\delta S_q = 0$ and introduce the Lagrange multipliers $\alpha$ and $\beta$ into the following equation,

$$\delta S_q - \alpha \delta M - \beta \delta E = 0, \tag{8}$$

which equals to

$$\int \left\{\frac{1}{1-q}(qf^{q-1}-1) - \alpha m - \frac{1}{2}\beta q f^{q-1} mv^2\right\}\delta f d^3x d^3v = 0. \tag{9}$$

Because Eq. (9) is satisfied independent on the choice $\delta f$, we have

$$\frac{1}{1-q}\left(qf^{q-1}-1\right)-\alpha m-\beta qf^{q-1}\frac{1}{2}mv^2=0. \tag{10}$$

And thus it determines the distribution function,

$$f=B_q\left[1-(1-q)\frac{\beta mv^2}{2}\right]^{1/1-q}. \tag{11}$$

Taking into account $\beta=1/(k_BT)$ and the normalization condition $\int fd^3x=n$, we find that the above distribution function is the same as the generalized velocity distribution (3). Thus, we realize that the generalized distribution function (3) is the extremum solution of Tsallis entropy. However, we are not sure whether this solution is stable or not. We have to seek for $\delta^2 S_q$ because, according to the stability theory, the stability is determined through the signs of the second variation of Tsallis entropy.

Then, we substitute (3) into (1) and take $T$ and $n$ as the variables to calculate the second variation of $S_q$ with respect to the solution (3). We obtain

$$\delta^2 S_q=\int\left[\frac{5-3q}{2}\frac{3}{2}\left(\frac{B_q}{B_0}\right)^{q-1}\frac{1}{T}\delta n\delta T+\frac{5-3q}{2}\frac{3}{2}\frac{1-3q}{4}\left(\frac{B_q}{B_0}\right)^{q-1}\frac{n}{T^2}(\delta T)^2\right]d^3x \tag{12}$$

By virtue of the mass and the energy conservation,

$$\delta M=\int \delta n d^3x=0, \tag{13}$$

$$\delta E=\delta\int\left[\delta n B_q^{q-1}+n\frac{5-3q}{2}\frac{B_q^{q-1}}{T}\delta T+\frac{5-3q}{2}\frac{B_q^{q-1}}{T}\delta n\delta T+n\frac{5-3q}{4}\frac{3(1-q)}{2}\frac{B_q^{q-1}}{T^2}(\delta T)^2\right]d^3x=0, \tag{14}$$

we can get a relation between $\delta n$ and $\delta T$ by

$$\delta n=-\left(n+n\frac{3}{4}(1-q)\frac{\delta T}{T}\right). \tag{15}$$

Substitute (15) into Eq.(12) to eliminate $\delta n$, it becomes

$$\delta^2 S_q=\int -\frac{3}{4}\frac{5-3q}{2}\left(\frac{B_q}{B_0}\right)^{q-1}\frac{n}{T^2}(\delta T)^2 d^3x. \tag{16}$$

Obviously, the signs of $\delta^2 S_q$ can be determined only by values of the nonextensive parameter $q$. We therefore obtain the results:

(i) If $q>5/3$, we have $\delta^2 S_q>0$ and the gas is unstable. As compared with (5) we find the instability of the gas in the nonextensive statistics is equivalent to the presence of negative specific heat. Although the negative specific heat is allowed for $q>5/3$ in the nonextensive statistics, it is still as a peculiarity of the *nonextensive* gas because the instability takes place. Thus, the existence of negative specific heat is explained reasonably from the stability analysis of the gas.

(ii) If $q < 5/3$, we have $\delta^2 S_q < 0$ and the gas is stable, corresponding to the case the specific heat is positive, $C_V > 0$. So, system is stable only if the parameter $q$ is bounded in the range of $q < 5/3$.

Now we see that the correspondence between the positive/negative sign of specific heat and the stability/instability conditions of the gas is established in nonextensive statistics. Meanwhile, we find that the quasi-equilibrium distribution of the nonextensive gas is not always stable for any $q$, the stability conditions depends on the value of $q$, and so the investigation of the connections between the values of $q$-parameter and the stability conditions of other systems may be important in the nonextensive statistics. It is worth to mention that thermodynamic stability problem was discussed in some works, e.g. Ramshaw [38] and Tsallis [39] considered thermodynamic stability with respect to the energy fluctuations by using the pairs {$S_q$, $U_1$} and {$S_q$, $U_q$}, respectively; Silva et al [40] showed that the specific heat was non-negative for the nonextensive parameter $q \notin [0,1)$ and associated with the thermodynamic stability of $q$-canonical ensemble by using the pair {$F_q$, $T$}. In present paper, the thermodynamic stability is studied based on the second variation of Tsallis entropy by using {$S_q$, $T$, $n$}. We show that the system is thermodynamically unstable only if the nonextensive parameter is $q>5/3$, which is exactly equivalent to the condition of appearance of the negative specific heat.

In conclusion, we have developed a technique of stability analysis of the classical ideal gas in the framework of nonextensive statistics and use this technique for the explanation of the negative specific heat. The stability analysis is made on the basis of the second variation of Tsallis entropy. It is shown that the system is unstable only if the nonextensive parameter is $q>5/3$, which is exactly equivalent to the condition of appearance of the negative specific heat. The connection of the stability of the nonextensive ideal gas with the sign of the specific heat and therefore with the values of the nonextensive parameter $q$ has been established.

**Acknowledgements**

We would like to thank the National Natural Science Foundation of China under the grant No.10675088 for the financial support.